# A NEW FULL ADDER CELL FOR MOLECULAR ELECTRONICS

Mehdi Ghasemi[1,2], Mohammad Hossein Moaiyeri[1,2], and Keivan Navi[1,2]

[1]Faculty of Electrical and Computer Engineering, Shahid Beheshti University G.C., Tehran, Iran.
[2]Nanotechnology and Quantum Computing Lab, Shahid Beheshti University G.C., Tehran, Iran.
navi@sbu.ac.ir

### *ABSTRACT*

*Due to high power consumption and difficulties with minimizing the CMOS transistor size, molecular electronics has been introduced as an emerging technology. Further, there have been noticeable advances in fabrication of molecular wires and switches and also molecular diodes can be used for designing different logic circuits. Considering this novel technology, we use molecules as the active components of the circuit, for transporting electric charge. In this paper, a full adder cell based on molecular electronics is presented. This full adder is consisted of resonant tunneling diodes and transistors which are implemented via molecular electronics. The area occupied by this kind of full adder would be much times smaller than the conventional designs and it can be used as the building block of more complex molecular arithmetic circuits.*

## 1. INTRODUCTION

The ability to use single molecules functioning as electronic devices has motivated researchers to minimize the size of the circuits in semiconductor industry. There have been advances previously such as microelectronics that has minimized the sizes through laws of physics. However, because of quantum mechanics and limitations of fabrication methods, further improvements could not be achieved. Packaging millions of silicon devices in a chip will lead to huge power consumption and expense [1]. Molecular electronics gives us flexibility in design such that we don't have it in traditional inorganic electronic materials. New molecular-electronic systems, architectures and analytical tools have been explored by chemists, physicists and engineers. Molecular components such as switches, rectifiers, transistors and memories are now arising. In fact, molecular electronics systems are composed of molecules with specific functions [2-5].

Molecular electronics has some advantages over other technologies for implementing logic circuits. The size of molecules is at least 1 nm and at most 100nm. This leads to less area and power dissipation. Switching can be done on the single-molecule scale. Many molecules have different geometric structures allowing distinct optical and electrical features. In addition, tools have been developed for molecular synthesis. There are also some disadvantages like instability at high temperature, reproducibility and the effective control of single-molecule transport [1].

There are two main molecular electronic structures utilized for logic design: polyphenylene-based chains and carbon nanotubes. However, it is much easier to design more complex logic structures using polyphenylene [6].

Polyphenylene chain has been used as resonant tunneling diodes, resistors and wires. Carbon nanotubes have been used in implementing logic and arithmetic circuits such as full adders [7,8], FPGA switches [9], multiple-valued logic circuits [10,11,12], analog and mixed-mode circuits [13] and CNFET-based crossbar memory [14]. Other molecular implementations of memories, latches, and switches have been reported in [15,16].

However, considering these circuits, full adder cell could be more of an interest due to its extensive usage in designing arithmetic circuits [17,18], which are usually located on the critical path of the VLSI systems such as microprocessors. Attempts have been made to reduce the area and enhance the performance of these cells. In this paper, we aim to represent a novel molecular full adder cell based on resonant tunneling diodes and molecular transistors, whereas the previous designs have been implemented just via resonant tunneling diodes. The proposed design leads to less complex and smaller design for the molecular full adder cell.

The rest of the paper is organized as follows: Section 2 discusses the polyphenylene-based components. Section 3 talks about the previous work. Section 4 presents the new full adder cell, and finally section 5 concludes the paper.

## 2. MOLECULAR COMPONENTS

In this paper we use polyphenylene-based components. To gain a special functionality in logic design, we should first build molecular structures as switches and then combine them into a complex circuit. We will discuss conductors, rectifying diodes, resonant tunneling diodes and transistors in this section.

### 2.1 Conductors

Polyphenylene-based conductors are composed of benzene rings with one or two hydrogen atoms eliminated. Benzene ($C_6H_6$) and its equivalent notations are illustrated in Figure 1 [19].

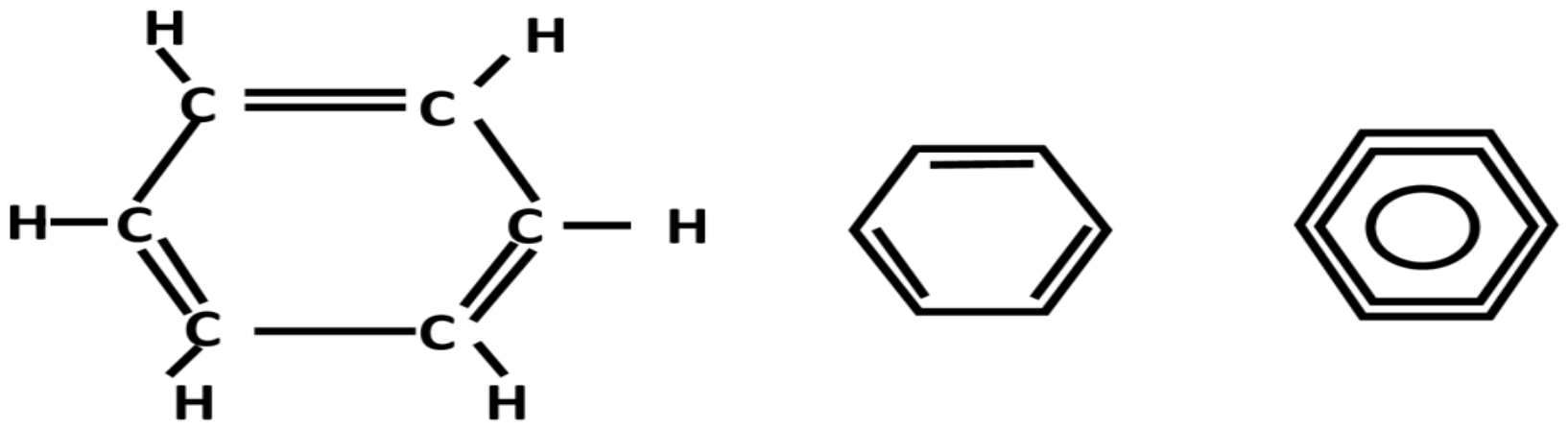

Figure 1. Benzene ring

In Figure 2, we see the phenyl and phenylene rings followed by a polyphenylene chain [19]. We get polyphenylene by binding phenylene to each other on both sides and terminating the result with phenyl groups. They can be made in different shapes and lengths. Benzene and polyphenylene molecules are called aromatic [5].

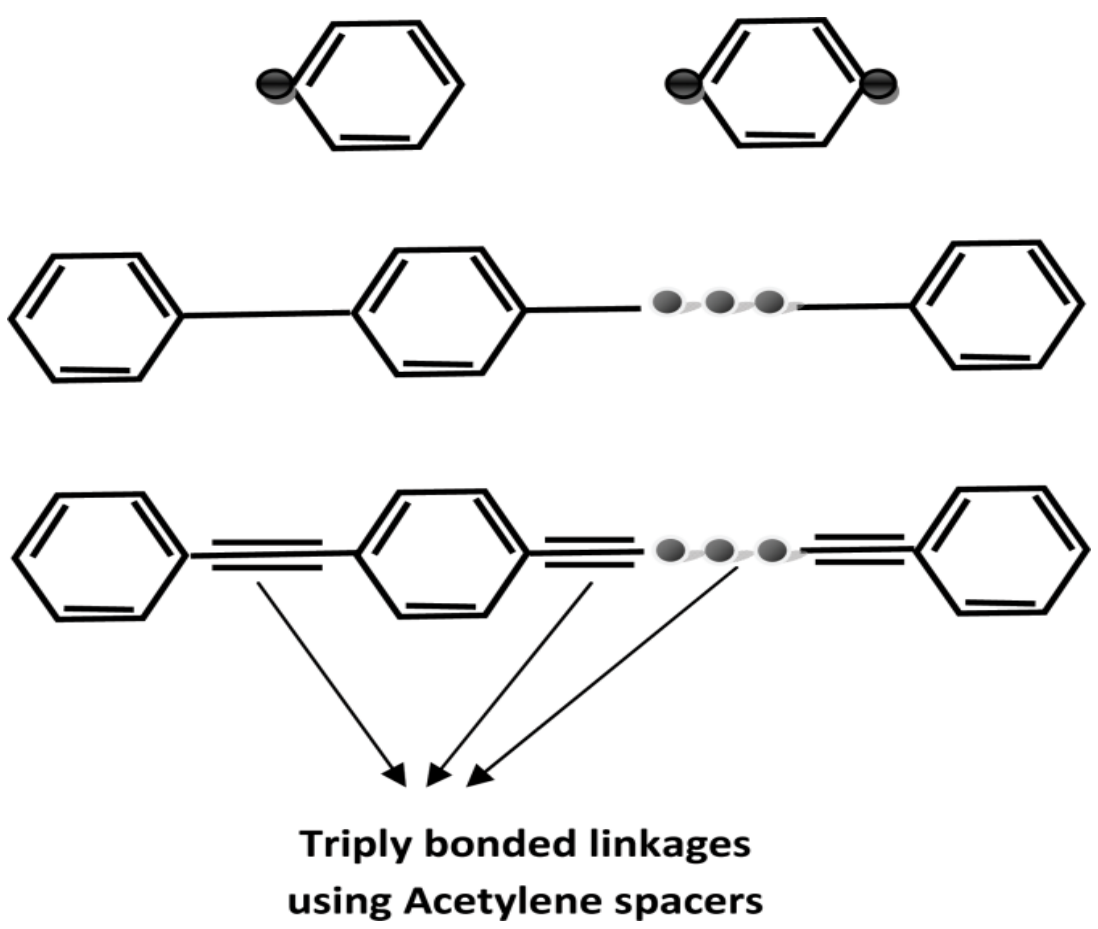

Figure 2. Polyphenylene chain

## 2.2 Diodes

Two types of diodes have been implemented via molecular electronics: rectifying diode, and resonant tunneling diode (RTD). In this section, we explain their molecular structures and functions.

### 2.2.1 Rectifying Diode

A rectifying diode is an electrical element which passes electric current through one direction. The direction which passes the current can be understood via the graphical symbol. It has a lot of applications in designing electric circuits. The first research about molecular electronics was based on rectifying diodes in 1974. The attempts showed that the lowest unoccupied molecular orbital (LUMO) and the highest occupied molecular orbital (HUMO) can be aligned in such a way that conduction is only feasible in one direction. In this way, constructing molecular diode is possible [6]. Figure 3 shows the structure of a rectifying diode. It consists of two sections S1, and S2 which is separated by an insulating group R [5,19].

Figure 4 depicts the schematic representation of a molecular rectifying diode [19].

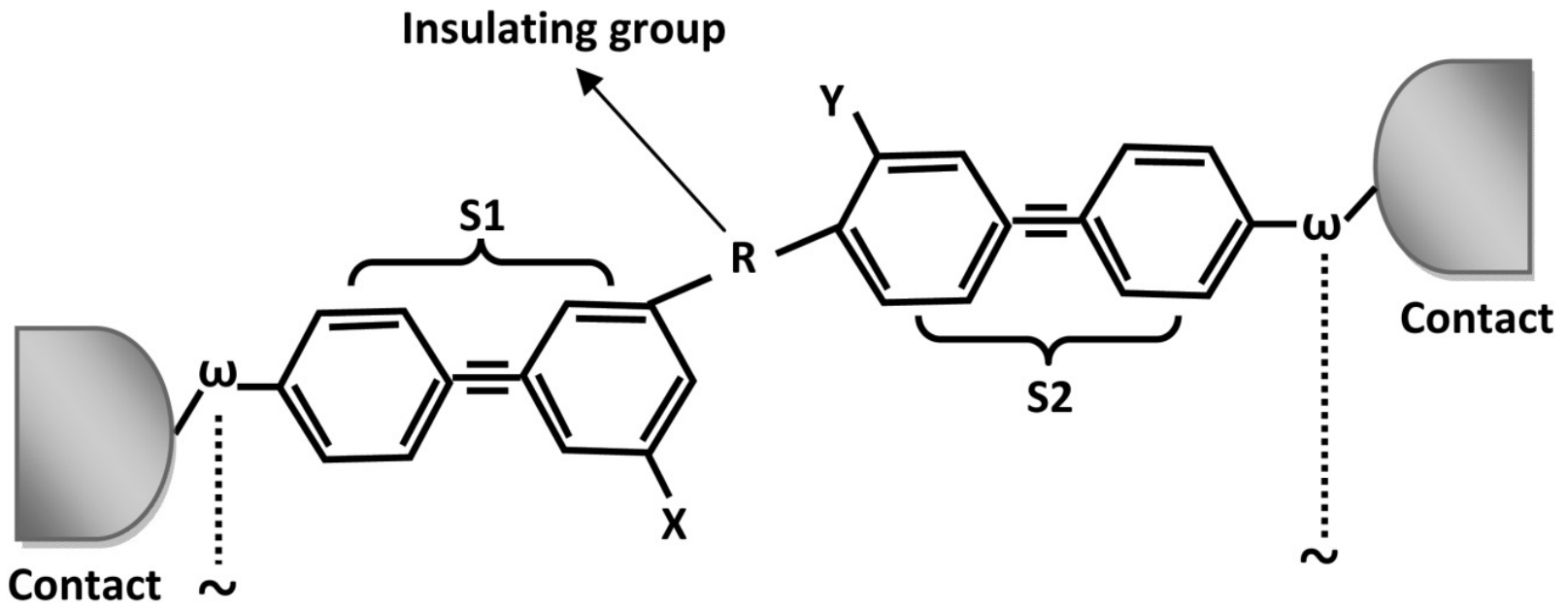

Figure 3. Structure of a molecular diode

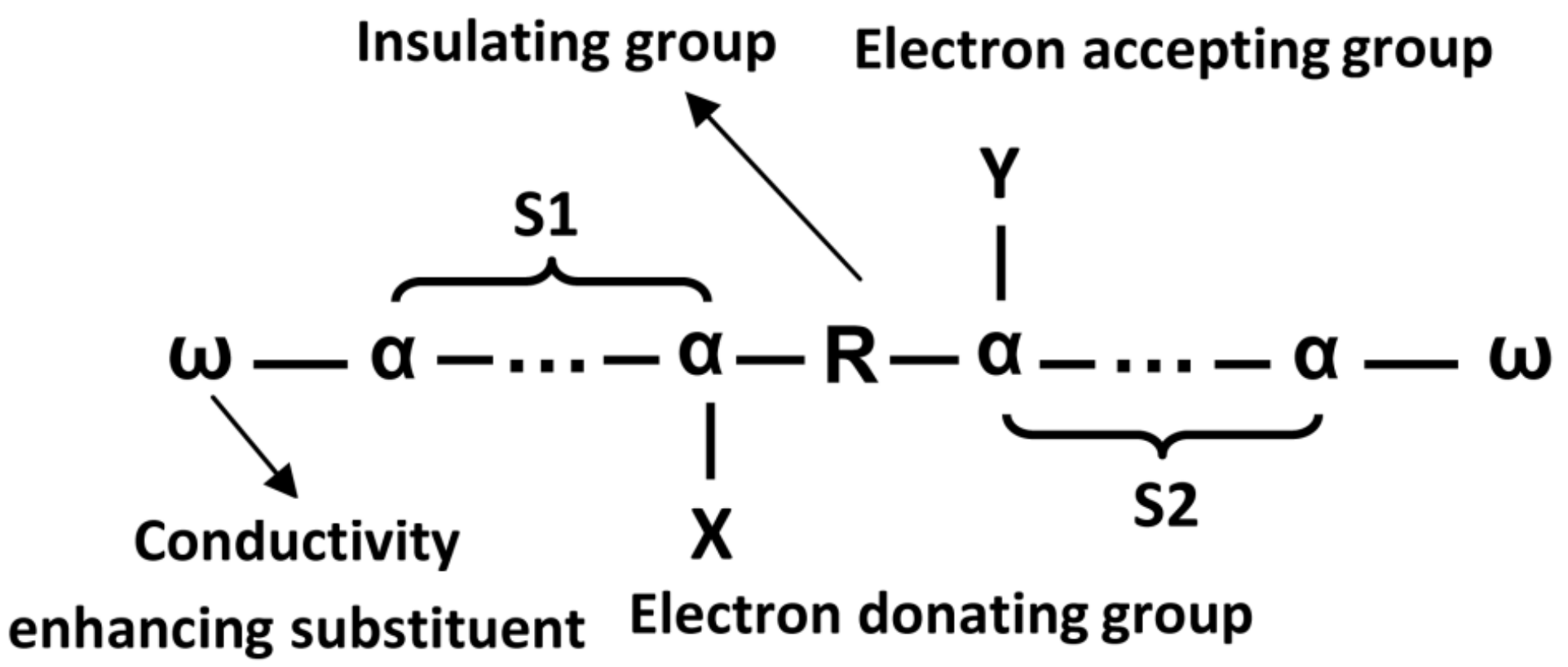

Figure 4. Schematic representation of a molecular diode

The band diagrams of this kind of rectifying diode are illustrated for zero bias, forward bias and reverse bias in Figure 5, Figure 6, and Figure 7, respectively [19]. As depicted, there are three barriers. Two of them are donor and acceptor sides and one of them is the insulating group. In Figure 5 $E_F$ refers to Fermi energy level. Electrons must have adequate energy to pass through barrier between two sections.

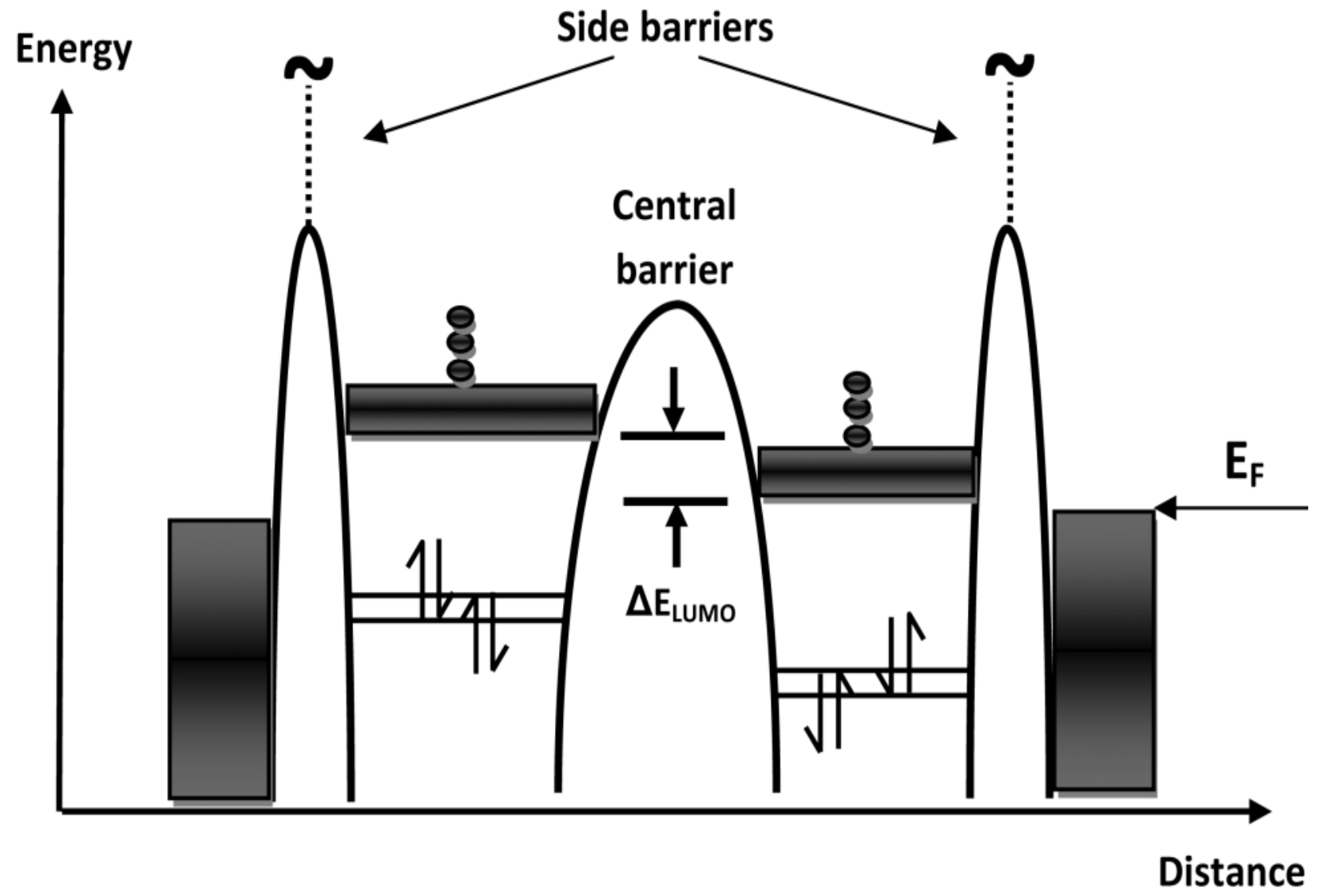

Figure 5. Energy level of a rectifying diode in the case of zero bias

In the case of forward bias, electrons move from right to left. Therefore under such circumstances, electric current flow from left to right.

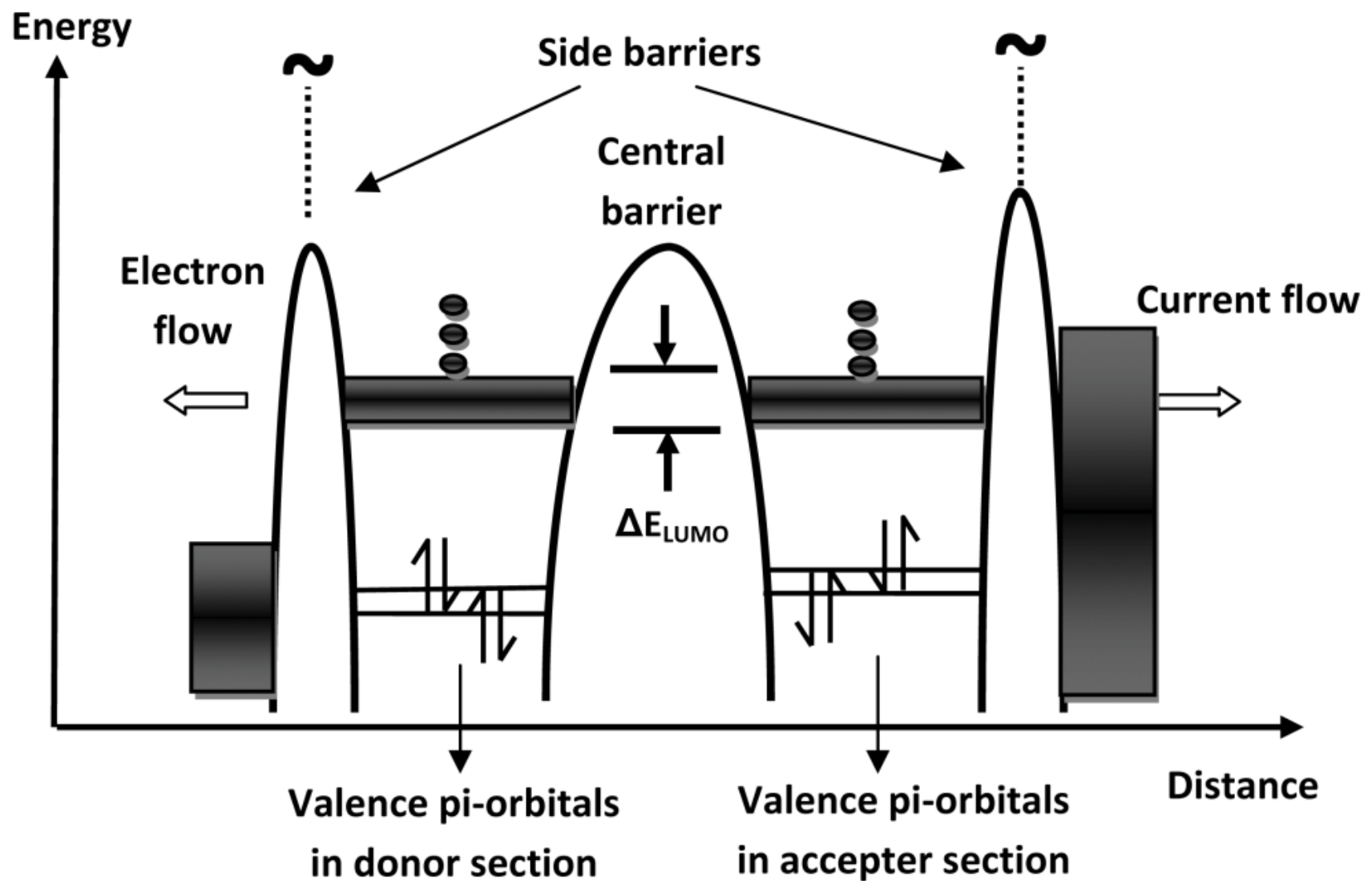

Figure 6. Energy level in the case of forward bias

In the last case and under reverse bias, the right side has a higher potential and electron flow is not feasible. Also, there is an energy difference between the $E_F$ of the left side and the LUMO energy of the right side.

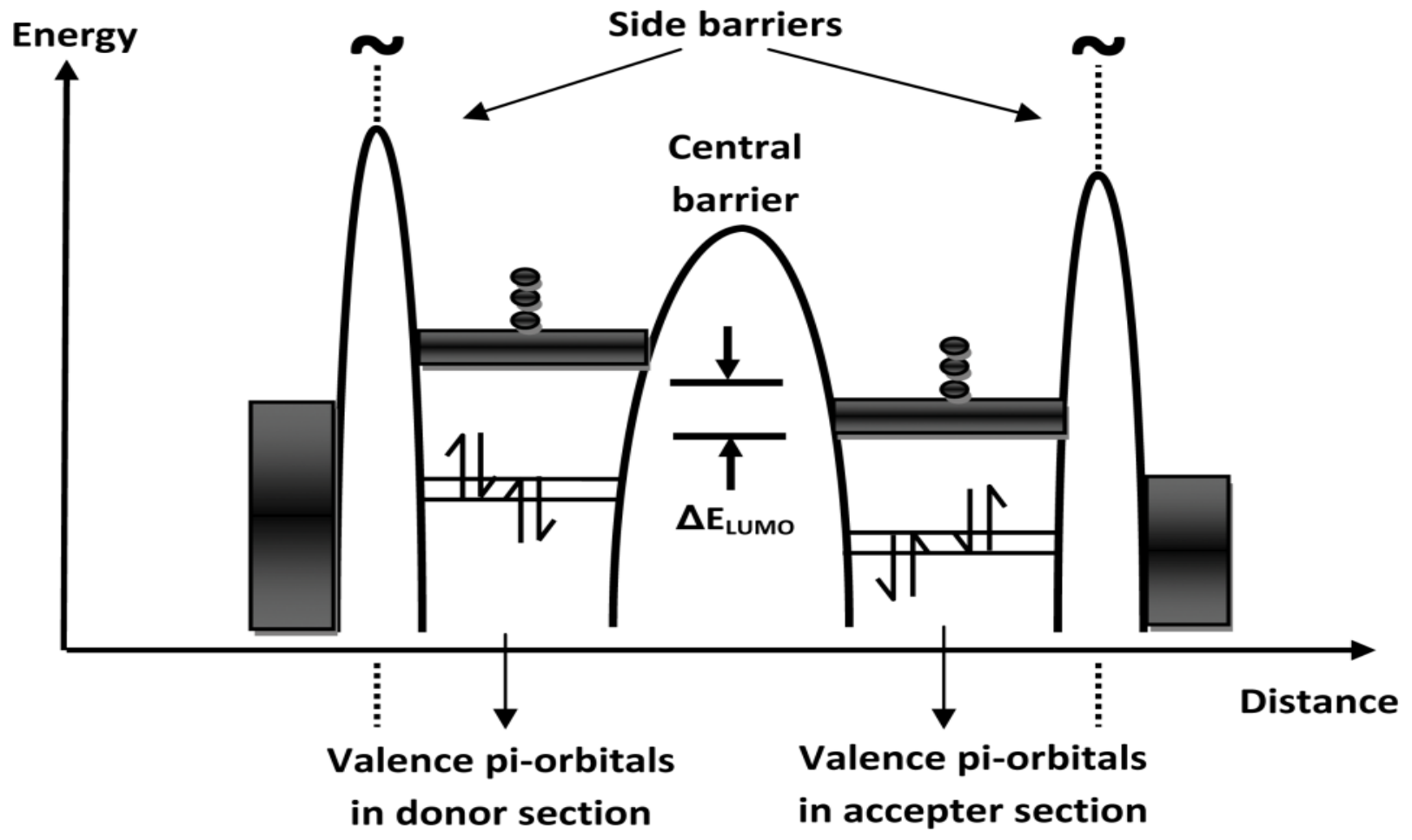

Figure 7. Energy level in the case of reverse bias

### 2.2.2 Resonant tunneling diode (RTD)

A resonant tunneling diode is a fast device with I-V characteristics shown in Figure 8 [20]. This device has numerous applications in the design of threshold logic circuits [21]. Instances of logic design such as AND, OR and majority function have been implemented using RTD together with transistors. Such designs can be evaluated through SPICE simulation.

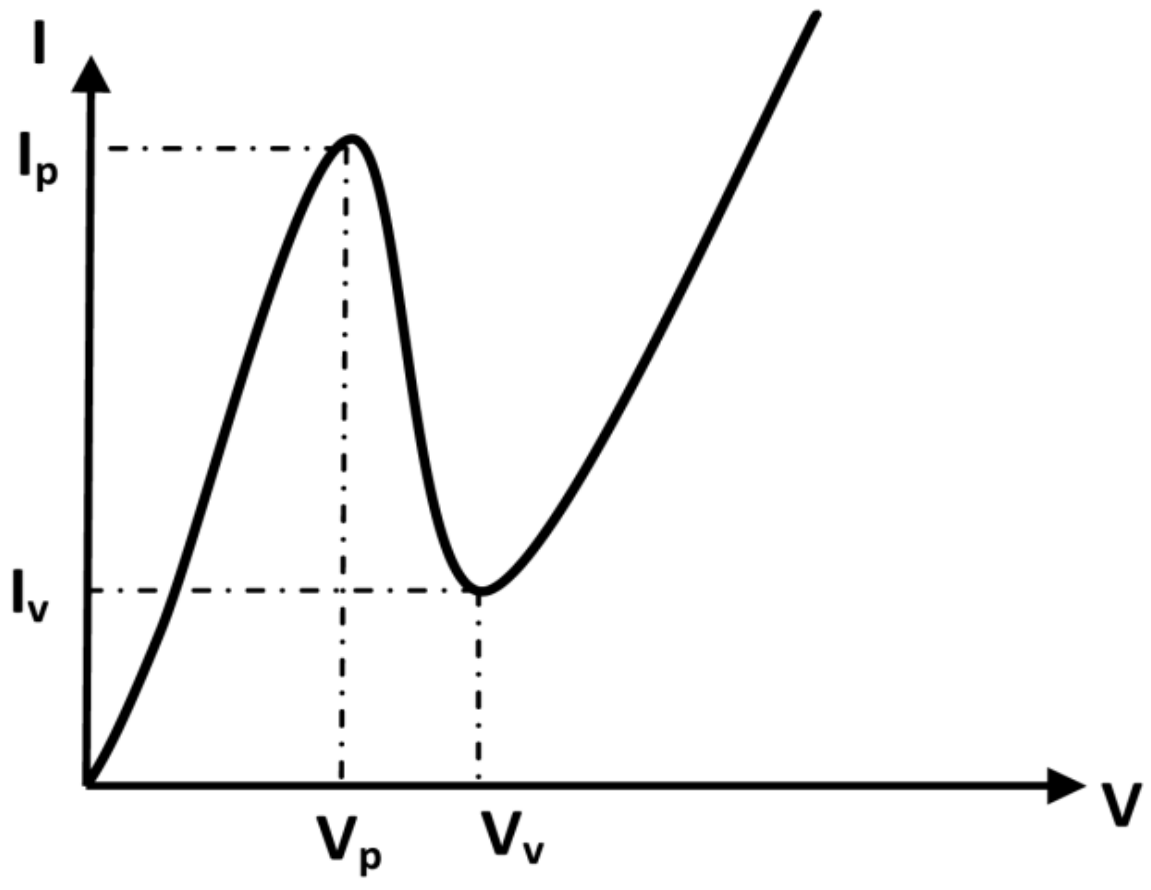

Figure 8. RTD I-V characteristics

Advances toward molecular electronics led to implementing molecular-based resonant tunneling diode [22,23]. The structure of a resonant tunneling diode based on molecular technology is shown in Figure 9 [24]. As we see in the figure below, there are two electron donating and accepting groups which are connected via $CH_2$ which acts as the barrier. This structure will provide us the aforementioned I-V characteristics.

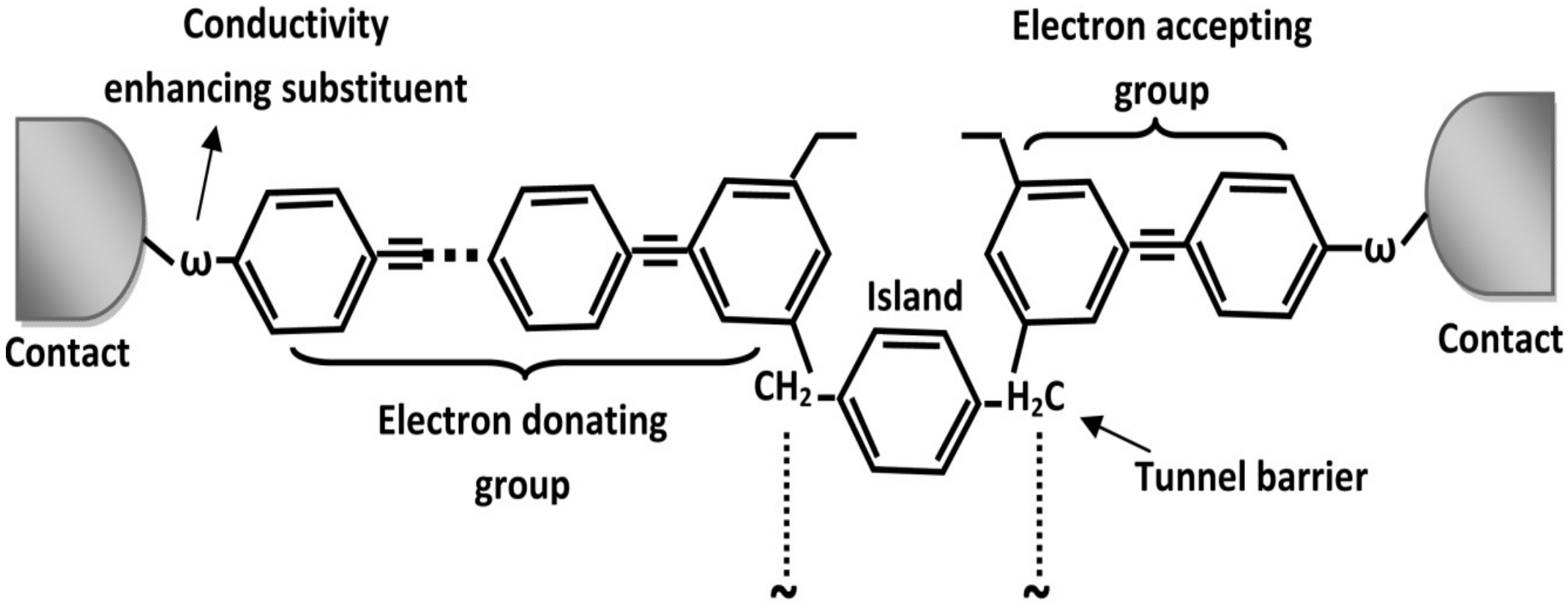

Figure 9. Resonant tunneling diode structure

## 2.3. Transistor

The benzene molecule can also play the role of a three-terminal device in such a way that the two electrodes attached act as the drain and source terminals and the other terminal is the capacitive gate [25,26]. This molecular device is called benzene-1, 4-dithiolate molecule. By replacing two opposite hydrogen atoms of the benzene ring with S atoms, such molecular device is formed. The sulphur atoms are then connected to gold electrodes.

This three terminal device is illustrated in Figure 10 [25]. The mechanism of such a device is also more described in [27,28]

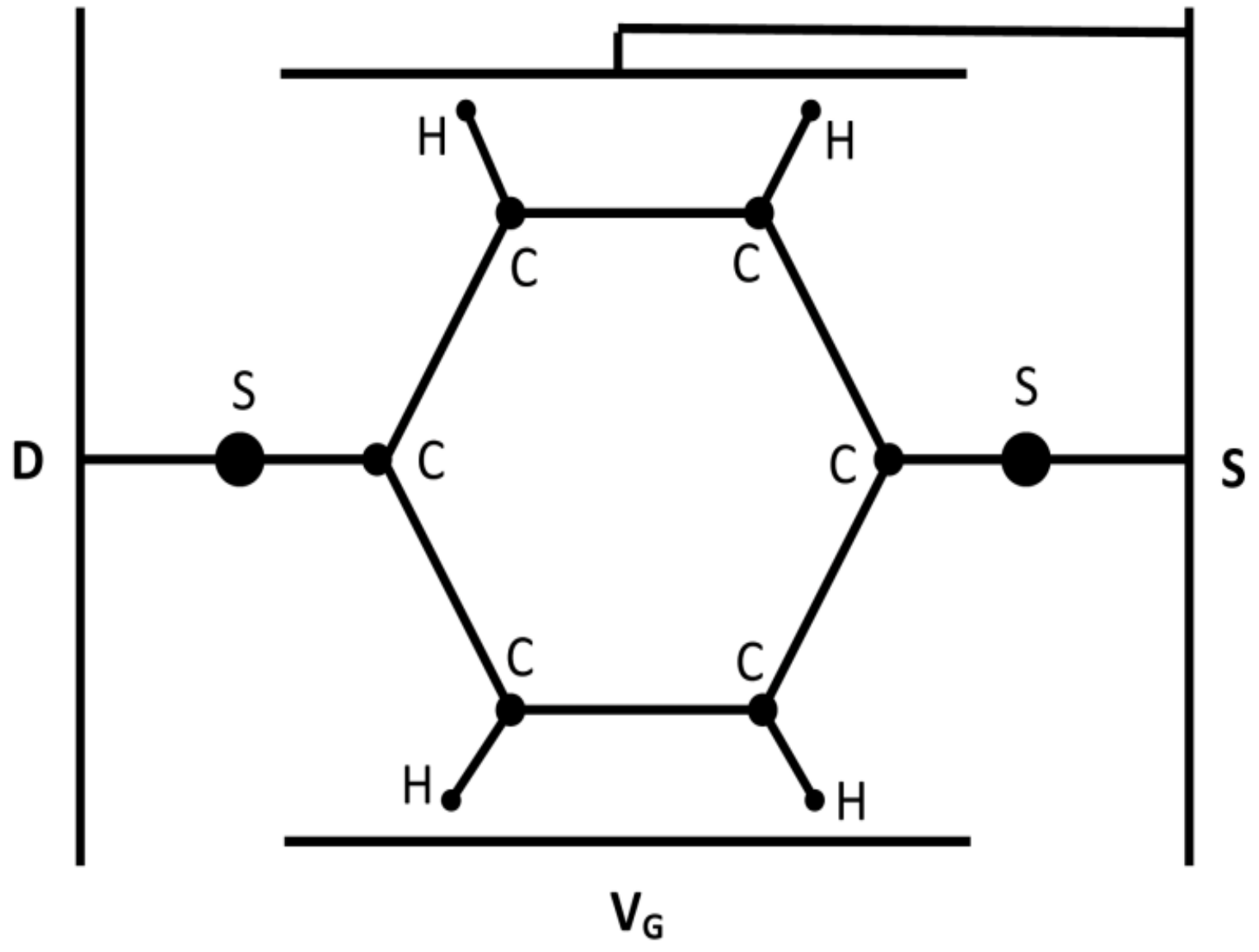

Figure 10. Structure of a three terminal transistor

## 3. Previous Work

In this section, we are going to review the previous efforts that have been made to design molecular adders. The basic logic gates like AND, OR, and XOR have been designed based on molecular diodes and other reconfigurable architectures [6,29]. Different other basic gates such as NOT, NAND and NOR can be constructed with the aid of aforementioned gates. The first two gates use two rectifying diodes and the last one uses one additional resonant tunneling diode [6]. These traditional designs merely use diodes and there is no usage of transistors.

Figure 11 is an example of a molecular half adder based on the mentioned scheme [30]. Another kind of molecular half adder is presented in [31] which is based on other types of molecular electronic devices. Using molecular transistors can considerably reduce the complexity of the design.

For designing full adders, the above schematic can be extended to achieve the goal [6]. However, designing the full adder cell based on half adder also increases the complexity of the design. We can see another instance of molecular full adder with different chemical combinations used [32].

There are also some programming techniques for designing logic circuits based on RTD [33,34]. A basic architecture for molecular programming has been proposed in [34] and gives us the idea of systematic way of designing. The final destination of molecular programming is to provide information processing systems at molecular levels.

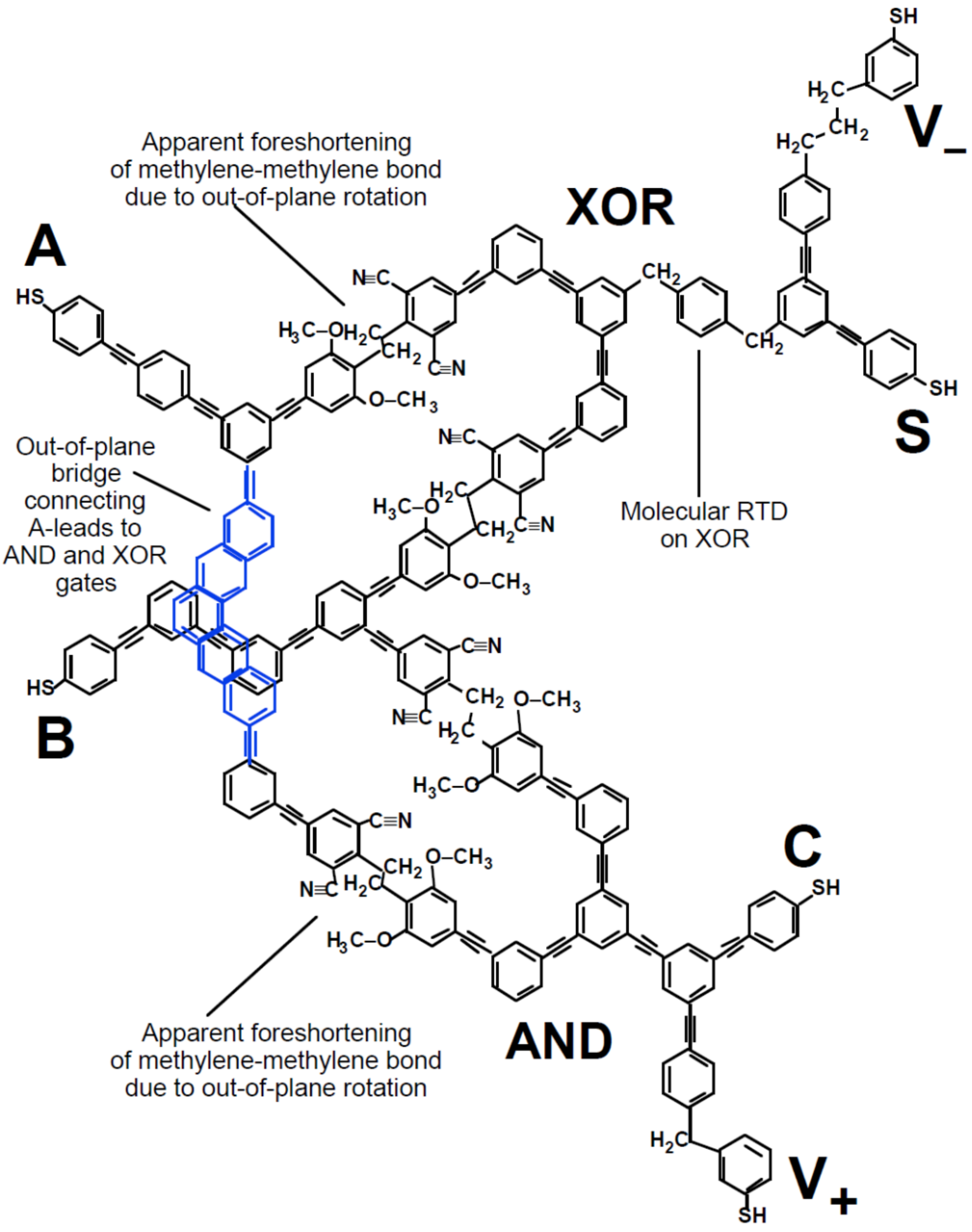

Figure 11. Molecular half adder [30]

## 4. THE PROPOSED FULL ADDER CELL

In this section we are going to present the new molecular full adder cell. In the literature, full adder cells are conventionally constructed using RTDs and transistors, which are compatible with the molecular electronics. In general, we can use threshold logic to design full adder cells. Majority function that can produce carryout is an instance of such methodology. We can also develop programming techniques to build more complex logic circuits based on these devices.

Figure 12 demonstrates the schema of the new molecular full adder cell. Since the operation of such circuits is in a parallel manner, this full adder can be used to implement carry propagate and faster adders like carry look ahead adder [35].

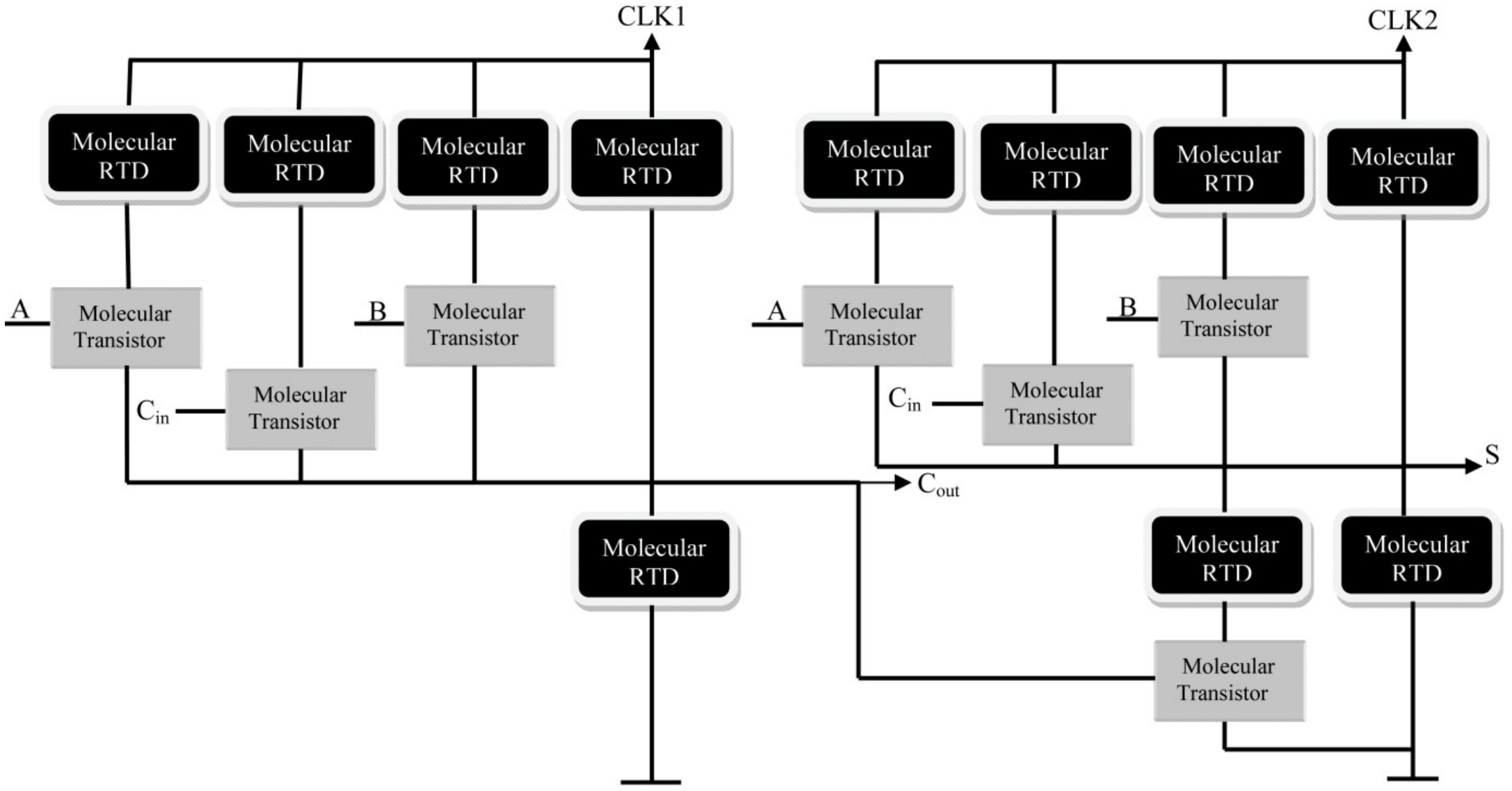

Figure 12. New molecular full adder

Figure 13 and Figure 14 demonstrate the molecular representations for the carry out and sum signals generator circuits of the proposed molecular full adder cell. In the proposed molecular designs, the inputs and outputs are implemented via SH contacts. Inputs to molecular transistors are perpendicular to the design plane. Other chemical combinations with different sizes can be used for such transistors. We can see that through molecular connections we can combine different molecular elements and build more sophisticated architectures. There are two clocks, one of them is used to generate carry out function and the other is utilized for sum function.

## 5. CONCLUSION

In this paper, fundamental issues about molecular electronics were introduced first and two kinds of molecular structures were described. Moreover, we investigated the molecular components involved in digital logic design and talked about previous designs.

The proposed design utilizes molecular three terminal devices and tunnelling diodes in implementing full adder cells. The area occupied by the proposed full adder cell is much times smaller than previous designs and reduces the complexity of the molecular Full Adder cell.
There have been great advances in the field of molecular transistors and future work can be done on designing threshold logic circuits based on these kinds of transistors. Future research can also be done on designing circuits having more regular structures and less area occupied. Other novel ideas may be using molecular majority function in designing full adders. There is the potential ability to use these structures in order to build more complex arithmetic structures and circuits. Likewise, great contributions can be done using hybridization techniques.

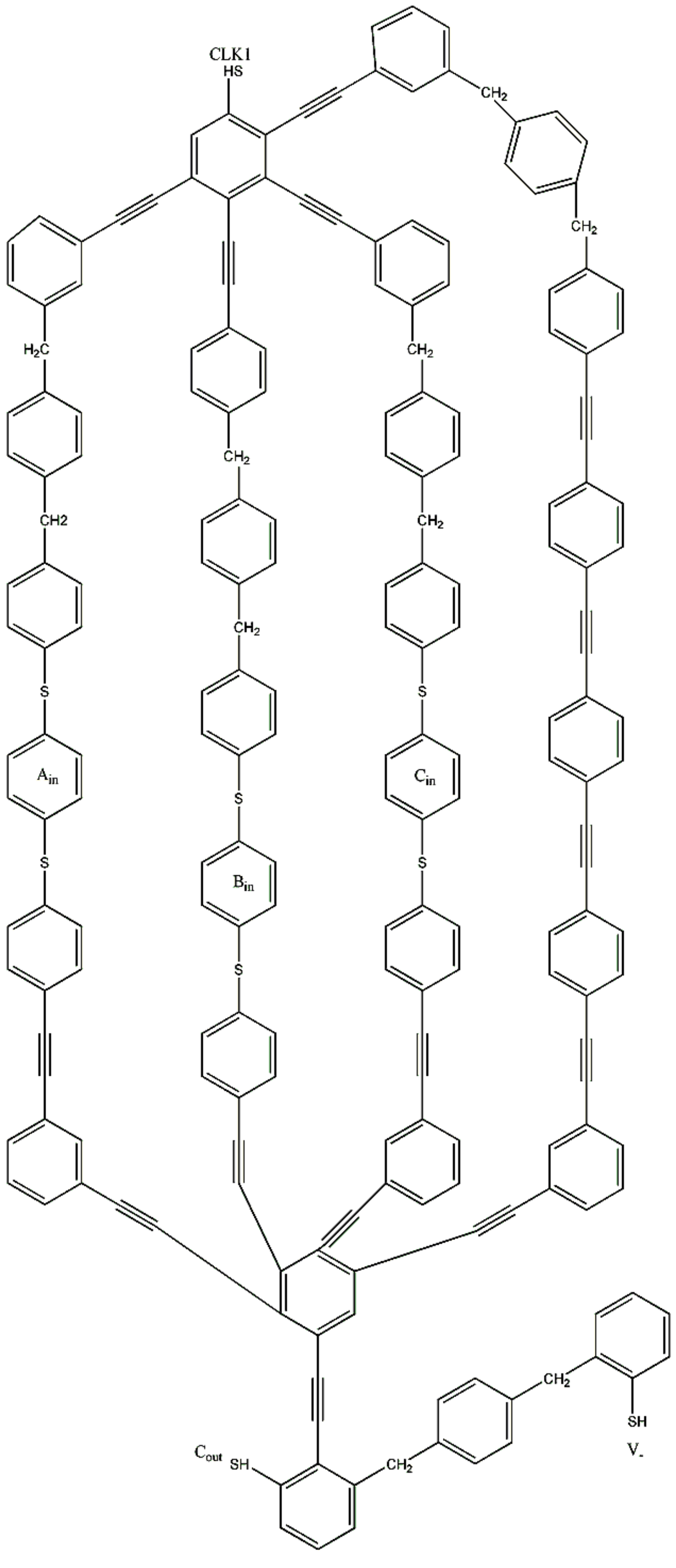

Figure 13. Molecular implementation of the output carry generator

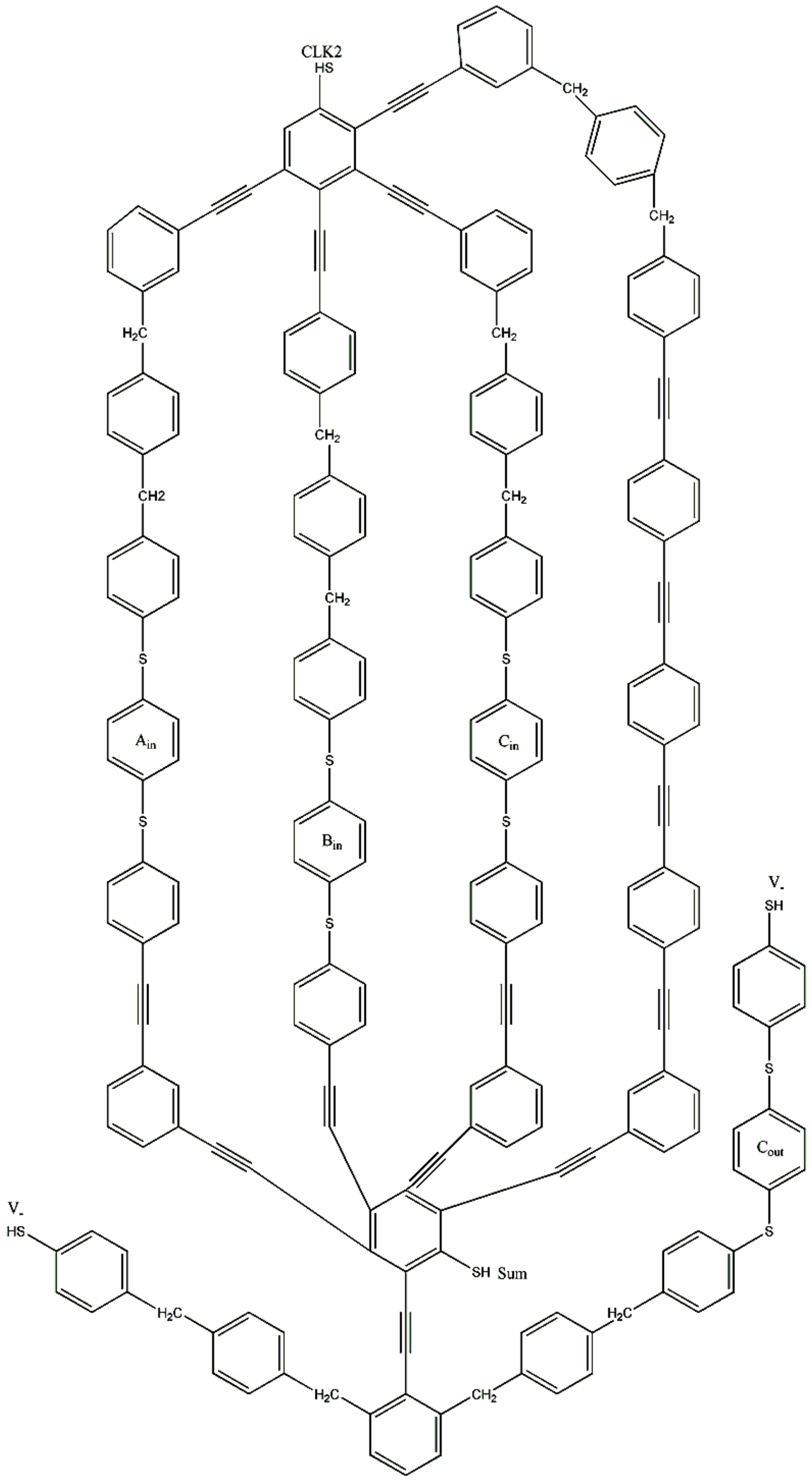

Figure 14. Molecular implementation of the sum generator